\documentclass[twocolumn]{aastex631}

\usepackage{hyperref}
\usepackage{mathrsfs}
\usepackage{amsmath}
\shorttitle{PTA High Redshifts}
\shortauthors{Wu et al.}
\newcommand{\customcite}[1]{\citeauthor{#1} \citeyear{#1}}

\begin{document}

\title{High-Redshift Merger Model for Low-Frequency Gravitational Wave Background}

\author[0000-0002-0196-9169]{Zhao-Feng Wu}
\email{Email: wu2177@purdue.edu} 
\author[0000-0003-1503-2446]{Dimitrios Giannios}
\affiliation{Department of Physics and Astronomy, Purdue University, 525 Northwestern Avenue, West Lafayette, IN 47907, USA}

\begin{abstract}
\noindent In 2023, the Pulsar Timing Array (PTA) Collaborations announced the discovery of a gravitational wave background (GWB), predominantly attributed to supermassive black hole binary (SMBHB) mergers. However, the detected GWB is several times stronger than the default value expected from galactic observations at low and moderate redshifts. Recent findings by the James Webb Space Telescope (JWST) have unveiled a substantial number of massive, high-redshift galaxies, suggesting more massive SMBHB mergers at these early epochs. Motivated by these findings, we propose an ``early merger" model that complements the standard merger statistics by incorporating these early, massive galaxies. We compare the early and standard ``late merger" models, which assume peak merger rates in the local Universe, and match both merger models to the currently detected GWB. Our analysis shows that the early merger model has a significantly lower detection probability for single binaries and predicts a $\sim 30 \%$ likelihood that the first detectable single source will be highly redshifted and remarkably massive with rapid frequency evolution. In contrast, the late merger model predicts a nearly monochromatic first source at low redshift. The future confirmation of an enhanced population of massive high-redshift galaxies and the detection of fast-evolving binaries would strongly support the early merger model, offering significant insights into the evolution of galaxies and SMBHs.
\end{abstract}


\section{Introduction} \label{sec:intro}

Last year, the North American Nanohertz Observatory for Gravitational waves (NANOGrav), the European PTA (EPTA), the Indian PTA (InPTA), the Australia-based Parkes PTA (PPTA), and the Chinese PTA reported evidence for the presence of a nanohertz GWB \citep{2023ApJ...951L...8A,2023A&A...678A..50E,2023ApJ...951L...6R,2023RAA....23g5024X}. The stochastic GWB is primarily expected to be the incoherent superposition of gravitational waves from SMBHBs across the entire Universe, although other possibilities exist \citep{2023ApJ...951L..11A}.

Current knowledge suggests that the properties of SMBHs are strongly correlated with those of their host galaxies \citep{2013ApJ...764..184M,2016ApJ...818...47S}, and the merger of SMBHs is likely to follow the merger of their host galaxies \citep{2004ApJ...615...19E,2013MNRAS.433L...1S}.
However, calculations of the GWB based on our local galactic astrophysical observations predict a value that is several times lower than the detected one \citep{2020ApJ...897...86C,2020MNRAS.498..537S,2022MNRAS.509.3488I,2023ApJ...952L..37A,2023MNRAS.524.4403M,2024MNRAS.528.1053C,2024A&A...685A..94E}. This discrepancy is not major, given the uncertainties in galactic observations, one can modestly adjust the modeling parameter values to bridge the gap between predictions and observations. 
\cite{2023ApJ...952L..37A} have shown that the observed GWB amplitude could be achieved with short binary hardening timescales, higher galaxy number densities, or higher normalization of the $M_{\mathrm{BH}}-M_{\text{bulge}}$ 
relation, resulting in more SMBH mergers, particularly involving more massive ones.

The unprecedented capabilities of JWST have revolutionized our understanding of the high-redshift Universe. So far, JWST has discovered a considerably large number of high-redshift massive galaxies, challenging our current understanding of galaxy formation and evolution \citep{2024arXiv240408052B,2024arXiv240416036U,2024ApJ...964L..10W}. The observed number of galaxies at the high-mass end of the galaxy stellar-mass function suggests a considerable number of massive galaxies already exist at $z \sim 5$ \citep{2024MNRAS.530..966G,2024arXiv240308872W}. JWST observations also indicate that SMBHs are much more massive than what the local $M_{\mathrm{BH}}-M_{\text{bulge}}$ relation expects at high redshift \citep{2023arXiv230801230M,2023ApJ...959...39H,2024ApJ...966..176Y,2023ApJ...957L...3P, 2024arXiv240405793M, 2024arXiv240303872J}.

Motivated by these new findings, we propose a new SMBHB merger population model consisting of two components. One component follows the standard merger model, with statistics reflecting the mean value of parameters from local galactic observations \citep{2014ApJ...783...85T}. The other component includes contributions from massive early galaxies, featuring a different type of population model and statistics, as well as a modified scaling relation at higher redshift. The new merger model aims to refine our understanding of the population of SMBHBs and their host galaxies, leveraging both GWB detected by PTAs and high-redshift data from JWST. We match both the proposed and standard models to the GWB detected by the current PTA and analyze the properties of the first detectable single binary predicted by these models. 

The structure of this paper is the following. In Section~\ref{2.1} and~\ref{2.2}, we model the distribution of number density of galaxy mergers and the subsequent SMBH mergers for the new and standard merger population models. We also construct the SMBHB population based on the merger rates in Section~\ref{2.3}. In Section~\ref{sec3}, we generate the GWB from the populations of SMBHBs and compare the GWB spectrum between different merger models. We study the properties of the first detectable single source, including the frequency evolution, detectability, mass, and redshift in Section~\ref{sec4}. In Section~\ref{motiva}, we present a more detailed motivation for proposing the new merger model and validate some of the assumptions in Section~\ref{validity}. We conclude in Section~\ref{sec:conclusion} with future outlooks.
Throughout this paper, we assume a WMAP9 cosmology, with $H_0=69.32$ km/s/Mpc, $\Omega_m = 0.2865$ and $\Omega_b = 0.04628$.

\section{Population Models} 

\begin{figure*}
    \centering
    \includegraphics[width=0.95\textwidth]{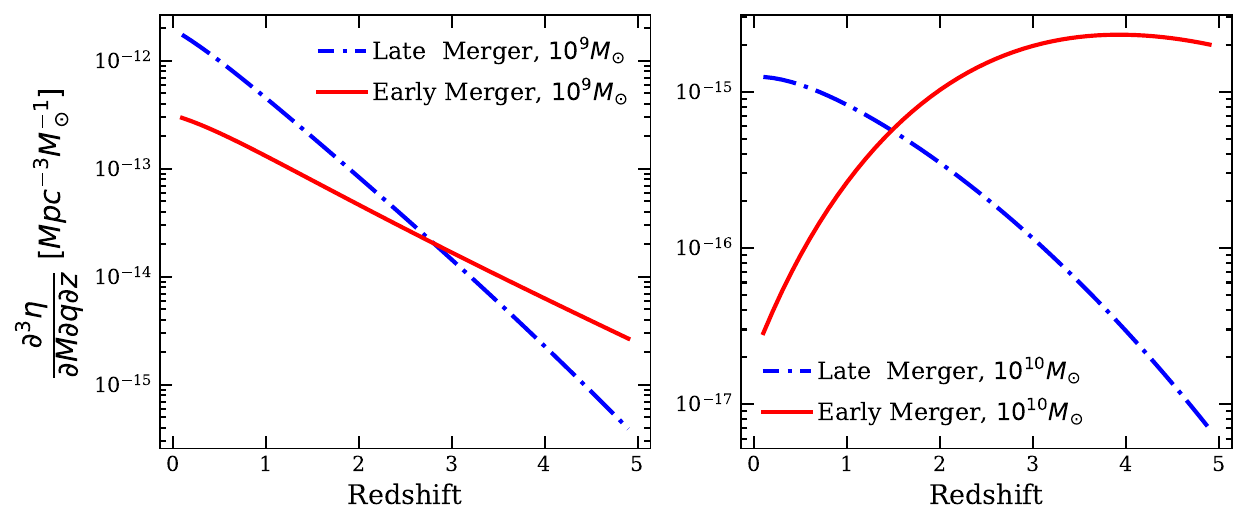}
    \caption{Comoving volumetric number density of binary mergers, $\eta(M,z)$, for binaries with equal mass ratios ($q=1$) across different total masses $M = 10^9 M_\odot$ (left plot) and $M = 10^{10} M_\odot$ (right plot). Contrary to the predictions of the standard `late merger' model, our proposed `early merger' model predicts a significantly higher merger density at greater redshifts for massive mergers. The `early merger' scenario assumes a substantially lower merger number density at low redshift for less massive binaries. The two models predict very similar gravitational wave background.}
    \label{fig:Merger_density}
\end{figure*}
Our objective is to investigate the characteristics of the SMBHB population capable of generating a GWB consistent with the NANOGrav, EPTA+InPTA, and PPTA recent results. We initiate our analysis by synthesizing the population of SMBHs that already merged in the cosmic history of the Universe.  

The currently detected GWB signal is several times stronger than the reference values suggested by local galactic observations \citep{2020ApJ...897...86C,2020MNRAS.498..537S,2022MNRAS.509.3488I,2023ApJ...952L..37A,2023MNRAS.524.4403M,2024MNRAS.528.1053C,2024A&A...685A..94E}. In this work, we explore two distinct scenarios to mitigate the gap between the prediction and observations. 

First scenario: we adjust the parameters from the standard expectations moderately so that more mergers, involving more massive black holes take place. In this model, most of the mergers of the SMBHs took place in the recent past. We call this merger population model as ``late merger" model.
 
Second scenario: we add one additional component to the standard merger model, which is the contribution from massive early galaxies as motivated by the JWST observations (see Section~\ref{motiva} for the details), and leave the other component with the parametric values as expected from local galactic observations. The additional population of SMBHs experience high-redshift mergers. Therefore, this proposed scenario is referred to as the ``early merger" model. 

Fig~\ref{fig:Merger_density} summarizes the difference between the population of SMBH mergers in the two scenarios. Contrary to the predictions of the standard ``late merger" model, our proposed ``early merger" model predicts a significantly higher merger density at greater redshifts for massive mergers. 

To obtain the population of SMBH mergers, we assume that all the mergers of SMBHs follow from the mergers of their host galaxies. In general, only three key components are necessary for modeling the SMBH merger populations responsible for the GWB: (i) the number density of galaxy mergers (ii) SMBH masses based on a galaxy–host relationship, and (iii) the lifetime of black hole binaries. 

We fix the lifetime of binaries to be $\tau_{\rm tot} = 0.1$ Gyr for all the mergers. With this simplification, we do not need to consider the evolution details of the binaries. See the Section~\ref{validity} for more discussion on the effects of this simplification. 

\subsection{Number Density of Galaxy Mergers} \label{2.1}
The number density of galaxy mergers $\left(\eta_{\text {gal-gal}}\right)$ can be expressed in terms of a galaxy stellar-mass function (GSMF;$\Psi(M,z)$), galaxy pair fraction (GPF;$P$), and galaxy merger time (GMT;$\left.T_{\text {gal-gal}}\right)$ \citep{2019MNRAS.488..401C}:
\begin{equation} \label{merger_density_rela}
\frac{\partial^3 \eta_{\text {gal}- \text {gal}}}{\partial m_{\star 1} \partial q_{\star} \partial z}=\frac{\Psi\left(m_{\star 1}, z^{\prime}\right)}{m_{\star 1} \ln (10)} \frac{P\left(m_{\star 1}, q_{\star}, z^{\prime}\right)}{T_{\text {gal}- \text {gal}}\left(m_{\star 1}, q_{\star}, z^{\prime}\right)} \frac{\partial t}{\partial z^{\prime}}
\end{equation}
This distribution is calculated in terms of the stellar mass of the primary galaxy $m_{\star 1}$, the stellar mass ratio $\left(q_{\star}=m_{\star 2} / m_{\star 1} \leqslant 1\right)$, and the redshift $z$. Because the galaxy and the subsequent SMBH merger span a finite timescale $\left(T_{\text {gal-gal}} + \tau_{\rm tot} \right)$, we distinguish between the initial redshift at which a galaxy pair forms $\left(z^{\prime}=z^{\prime}[t]\right.$ at some initial time $\left.t\right)$ and the redshift at which the system becomes a postmerger remnant $\left(z=z\left[t+T_{\text {gal}-\text {gal}}+ \tau_{\rm tot} \right]\right)$. 

The GSMF is defined as
\begin{equation}
\Psi\left(m_{\star 1}, z^{\prime}\right) \equiv \frac{\partial \eta_{\star}\left(m_{\star 1}, z^{\prime}\right)}{\partial \log _{10} m_{\star 1}},
\end{equation}
i.e., the differential number density of galaxies per decade of stellar mass. There are different choices of the GSMF, a standard implementation \citep{2023ApJ...952L..37A,2019MNRAS.488..401C} describes the GSMF in terms of a single-Schechter function \citep{1976ApJ...203..297S},
\begin{equation} \label{eq3}
\Psi\left(m_{\star 1}, z\right)=\ln (10) \Psi_0 \cdot\left[\frac{m_{\star 1}}{M_\psi}\right]^{\alpha_\psi} \exp \left(-\frac{m_{\star 1}}{M_\psi}\right)
\end{equation}
where we have introduced $\Psi_0, M_\psi$, and $\alpha_\psi$ as new variables. Since the GSMF can vary with redshift, we parameterize these quantities as
\begin{equation} \label{para_eq}
\begin{aligned}
\log _{10}\left(\Psi_0 / \mathrm{Mpc}^{-3}\right) & =\psi_0+\psi_z \cdot z \\
\log _{10}\left(M_\psi / M_{\odot}\right) & =m_{\psi, 0}+m_{\psi, z} \cdot z \\
\alpha_\psi & =1+\alpha_{\psi, 0}+\alpha_{\psi, z} \cdot z
\end{aligned}
\end{equation}
such that each of these quantities has a simple linear scaling with redshift. This introduces six new dimensionless parameters into our models, corresponding to the normalization $\left(\psi_0, m_{\psi, 0}\right.$, and $\left.\alpha_{\psi, 0}\right)$ and slope $\left(\psi_z, m_{\psi, z}\right.$, and $\left.\alpha_{\psi, z}\right)$ of the redshift scaling. In all of the analyses presented here, only the GSMF normalization and characteristic mass parameters $\psi_0$ and $m_{\psi, 0}$ are allowed to vary, all the other parameters are kept fixed at the fiducial values specified in Table \ref{tab:model_components}.

There are two different choices of the parameters for $\psi_0$ and $m_{\psi, 0}$, one derived from local galaxy observation mean values \citep{2014ApJ...783...85T} and the other used results of the Phenom+Astro model from \cite{2023ApJ...952L..37A}. These sets of parameters correspond to the Galaxy Observation Mean Value (GOMV) model and the late merger model, respectively, and they are all described by only a single-Schechter function for each.

\begin{figure}
    \includegraphics[width=\linewidth]{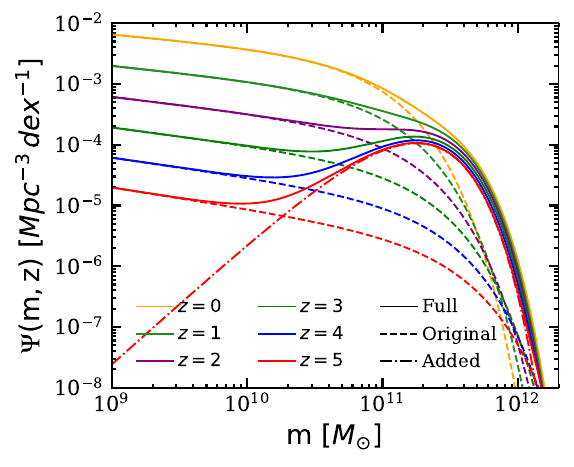}
    \caption{Comparison of GSMFs $\Psi(m,z)$ for standard and early merger models across redshift bins $0 < z < 5$. Dashed lines represent the GSMFs under single-Schechter functional form with parameters derived from observations at low and moderate redshifts \citep{2014ApJ...783...85T}. Dash-dot lines indicate a possibly additional population of massive galaxies that already exists at high redshifts. Solid lines show the combined GSMFs, integrating standard and additional massive galaxy populations. The divergence between the synthetic and standard models is most notable at the high-mass end and increases with redshift.}
    \label{fig:GSMF_dec}
\end{figure}

The early merger model assumes a more complicated GSMF. Figure~\ref{fig:GSMF_dec} shows the distributions characterized by the GOMV (Original) and the distribution of an extra component as we proposed (Added). The GSMFs of the early merger model are the sum of the original and added distributions. Equivalently, the early merger model describes the GSMF by a double-Schechter function, with an additional single-Schechter function describing the possible high-redshift massive component of the whole galaxy populations,
\begin{equation}
\begin{aligned}
\Psi \left(m_{\star 1}, z\right) & =\ln (10) \Psi_{0,a} \cdot\left[\frac{m_{\star 1}}{M_{\psi,a}}\right]^{\alpha_{\psi, a}} \exp \left(-\frac{m_{\star 1}}{M_{\psi,a}}\right) + \\
& \ln (10) \Psi_{0,o} \cdot\left[\frac{m_{\star 1}}{M_{\psi,o}}\right]^{\alpha_{\psi,o}} \exp \left(-\frac{m_{\star 1}}{M_{\psi,o}}\right)
\end{aligned}
\end{equation}
Here $\Psi_{0,a}$, $M_{\psi,a}$, $\alpha_{\psi,a}$, $\Psi_{0,o}$, $\alpha_{\psi,o}$, and $M_{\psi,o}$ are also defined in a similar way as Equation \ref{para_eq}. $\Psi_{0,o}$, $\alpha_{\psi,o}$, and $M_{\psi,o}$ denote the original component, with values following the same values as GOMV. $\Psi_{0,a}$, $\alpha_{\psi,a}$, and $M_{\psi,a}$ are parameters describing the newly added component in Figure~\ref{fig:GSMF_dec} with numerical values listed in Table \ref{tab:model_components}. The values are chosen to produce the observed GWB amplitude.

\begin{figure}
    \includegraphics[width=\linewidth]{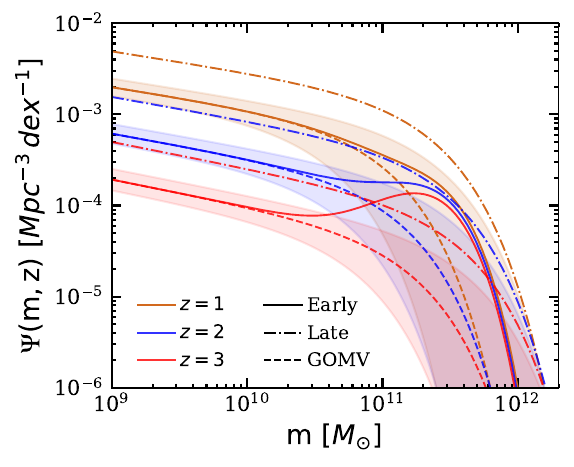}
    \caption{GSMF $\Psi(m,z)$ for various models across the redshift range $1 < z < 3$. Dashed lines represent the GSMF according to the GOMV model, solid lines show the GSMF of the early merger model, and dash-dot lines indicate the late merger model. The shaded areas denote uncertainties due to turnover mass variations, with the uncertainty range inferred from \cite{2014ApJ...783...85T}. The early merger model's GSMF closely matches observations up to $z=2$ but diverges notably at $z=3$, exhibiting a prominent excess at $m \sim 10^{11} M_\odot$. In contrast, the GSMF of the late merger model shows moderate deviations across all masses and redshifts.}
    \label{fig:GSMF_ob}
\end{figure}

Although we include an additional population of high-redshift massive galaxies, the new GSMF remains consistent with local galactic observations \citep{2014ApJ...783...85T} for $z \lesssim 2$. In Figure~\ref{fig:GSMF_ob}, the shaded areas indicate the uncertainties in the GSMFs. These uncertainties are based on the posterior width of the turnover mass $M_{\psi}$, as inferred from \cite{2014ApJ...783...85T} using the model described by Equations~\eqref{eq3} and~\eqref{para_eq}. We focus on the turnover mass since it dominates the critical high-mass end of the GSMFs.

The GSMF of the early merger model shows a prominent excess at $m \sim 10^{11} M_\odot$ for $z > 2$. In contrast, the GSMF of the late merger model has moderate deviations from the local galactic observations across all masses and redshifts. These discrepancies can be mitigated by adjusting either the scaling relation or the redshift dependence of the parameters to produce a similar GWB level. The early merger model is more flexible, as its two components can be tuned separately.

However, if we assume a GSMF and scaling relation within local observational error margins and extrapolate to higher redshifts, the resulting GWB is significantly lower than detected. The primary aim of this paper is to investigate the different predictions arising from two distinct models for the SMBHB merger history. Therefore, we do not focus on the specific parameter values within these models and leave detailed modeling for future studies as more observational data become available.

After modeling the GSMF, then come the GPF and GMT. The GPF describes the number of observable galaxy pairs relative to the number of all galaxies. The GMT is the duration over which two galaxies can be discernible as pairs from an initial separation at which they are associated with one another until a final separation, after which they are no longer distinguishable as separate galaxies. These two distributions are typically determined empirically based on the detection of galaxy pairs in observational surveys and thus depend on observational definitions and selection criteria \citep{2017MNRAS.468..207S,2019ApJ...876..110D}

In practice, we parameterize $P\left(m_{\star 1}, q_{\star}, z^{\prime}\right)$ and $T_{\text {gal-gal }}\left(m_{\star 1}, q_{\star}, z^{\prime}\right)$ as power laws of $q_{\star}$ and $z$ following \cite{2019MNRAS.488..401C}:
\begin{equation}
P\left(m_{\star 1}, q_{\star}, z^{\prime}\right) =P_0(1+z)
\end{equation}
\begin{equation} \label{galaxy_time}
T_{\text {gal-gal}}\left(m_{\star 1}, q_{\star}, z^{\prime}\right) =T_0(1+z)^{\beta_t} q^{\gamma_t} \\
\end{equation}
All the above parameters are listed in Table~\ref{tab:model_components} and kept fixed during our subsequent analyses.

\subsection{SMBH Merger Density} \label{2.2}
The next step is to connect the properties of the SMBHs to those of the host galaxies. In this work, we assume a one-to-one correspondence between the properties of galaxy pairs before the merger and the subsequent SMBH binaries. In other words, we adopt an SMBH-host relationship to translate from galaxies to SMBHs. In this analysis, we restrict ourselves to the $M_{\mathrm{BH}}-M_{\text{bulge}}$ relationship, which relates the galaxy stellar bulge mass to the SMBH mass for each component of the binary as \citep{2003ApJ...589L..21M}
\begin{equation}
\log _{10}\left(M_{\mathrm{BH}} / M_{\odot}\right)=\mu+\alpha_\mu \log _{10}\left(\frac{M_{\text {bulge }}}{10^{11} M_{\odot}}\right)
\end{equation}
Here we ignore the error term for simplicity and it will not change the conclusion of the paper. This relation depends on the dimensionless black hole mass normalization $\mu$ and power-law index $\alpha_\mu$. 

In light of recent observations on overmassive SMBH at high redshifts \citep{2010ApJ...709..937G,2023ApJ...950...68E,2023ApJ...957L...3P,2024Natur.628...57F}, we modify $\mu$ to have a strong redshift dependence for the added part of the early merger model, with higher redshift tending to have more massive SMBH. Specific choice of the value is also given in Table~\ref{tab:model_components} and more discussions in Section~\ref{motiva}.

Note that only a fraction of the galaxy stellar mass is in the stellar bulge component $\left(M_{\text {bulge}}=f_{\star, \text{bulge}} \cdot m_{\star}\right)$, which we take to be $f_{\star, \text{bulge}}=0.615$ based on empirical bulge fraction measurements of massive galaxies from \cite{Lang_2014} and \cite{10.1093/mnras/stu594}. In principle, the $f_{\star, \text{bulge}}$ could depend on the stellar mass \citep{2016MNRAS.457.1308M}, but it is also likely related to the redshift and galaxy type \citep{2024A&A...685A..48H,2024A&A...683L...4J}. However, in our study, $f_{\star, \text{bulge}}$ is degenerate with $\mu$ and $\alpha_\mu$, and none of these parameters are yet well understood or constrained at high redshifts. Therefore, to simplify the model, we fixed $f_{\star, \text{bulge}}$ as a constant throughout the paper, leaving more detailed treatments for future work.

Using the $M_{\mathrm{BH}}-M_{\text{bulge}}$ relationship, we transform the number density of galaxy mergers to a number density of SMBH binaries via
\begin{equation} \label{eta_merger}
\frac{\partial^3 \eta}{\partial M \partial q \partial z}=\frac{\partial^3 \eta_{\text{gal}-\text {gal}}}{\partial m_{\star 1} \partial q_{\star} \partial z} \frac{\partial m_{\star 1}}{\partial M}
\end{equation}
Here we approximate the mass ratio of the galaxies as the mass ratio of the SMBHs, as we ignore the intrinsic scatter of the $M_{\mathrm{BH}}-M_{\text{bulge}}$ relation. 

\subsection{SMBHB Population} \label{2.3}
In our study, we are also interested in the population of SMBHBs. Therefore, we define $d N\equiv \eta(M,q,z) d V_c$, which is the comoving number of binaries emitting in a given logarithmic frequency interval per unit total mass, mass ratio, and redshift \citep{2008MNRAS.390..192S}. Then by definition,
\begin{equation} \label{pop_from_density}
\frac{\partial^4 N}{\partial M \partial q \partial z \partial \ln f_p}=\frac{\partial^3 \eta}{\partial M \partial q \partial z} \frac{\partial t}{\partial \ln f_p} \frac{\partial z}{\partial t} \frac{\partial V_c}{\partial z}.
\end{equation}

Now we define the binary hardening timescale, $\tau\left(f_p\right) \equiv f_p /\left(d f_p / d t\right) = dt/dln f_p$, as the rest-frame duration a binary spends within a specific logarithmic frequency interval. Integrating this with the comoving volume element of a light cone,
\begin{equation}
\frac{\partial V_c}{\partial z} \frac{\partial z}{\partial t}=4 \pi c(1+z) d_c^2,
\end{equation}
where $d_c$ denotes the comoving distance at redshift $z$, yields
\begin{equation} \label{pop_to_density}
\frac{\partial^4 N}{\partial M \partial q \partial z \partial \ln f_p} = \frac{\partial^3 \eta}{\partial M \partial q \partial z} \cdot \tau\left(f_p\right) \cdot 4 \pi c(1+z) d_c^2.
\end{equation}

For circular binaries evolving solely through quadrupole GW emission, the semimajor axis's rate of change and the corresponding hardening timescale are described by
\begin{equation} \label{gw_time}
\tau_{\mathrm{gw}}(\mathcal{M}, f_p) \equiv \left(\frac{\partial t}{\partial \ln f_p}\right)_{\mathrm{gw}}=\frac{5}{96}\left(\frac{G \mathcal{M}}{c^3}\right)^{-5 / 3}\left(2 \pi f_p\right)^{-8 / 3}.
\end{equation}

Notice that the interactions of the binary with the environment are important for the decay of the orbits before the GW emission becomes dominant. These interactions are necessary for the binary to merge within the Hubble time \citep{2024A&A...685A..94E,2023ApJ...952L..37A}. However, taking into account the details of the interaction is too complicated and beyond the scope of this paper. Depending on the mechanism of interactions, the hardening timescale varies dramatically, so does the population of the binaries. However, we expect the first detected single binary by PTAs to have already entered the GW-dominated phase and is, therefore, insensitive to these interactions. More discussion on this point in Section~\ref{validity}.

Another problem is assuming a continuous distribution of binaries across the $(M, q, z, f)$ parameter space in Equation~(\ref{pop_to_density}). At low frequencies, the hardening timescale is very long, and a large number of binaries contribute to the GWB, making this approximation valid. At higher frequencies, however, the hardening timescale becomes shorter, and the typical number of binaries producing the bulk of the GWB energy in a given frequency bin approaches unity \citep{2008MNRAS.390..192S}. In this regime, a continuous distribution overestimates the GWB signal. Properly accounting for the finite number of sources in each frequency bin therefore results in a steeper GWB spectrum at high frequencies \citep{2008MNRAS.390..192S}. 

To discretize the SMBH binary population, we assume that the true number of binaries in any given spatial volume is Poisson-distributed. Notice that Equation~(\ref{pop_hc}) provides an expectation value for the number of binaries in a point $(M, q, z, f)$ in parameter space. We then integrate the differential number of binaries over finite bins of parameter space to obtain the expected number of binaries in each bin. Then the number of binaries in each bin is given by a Poisson distribution $(\mathcal{P})$
\begin{equation} \label{pop_gen}
    N_\mathcal{P}(M, q, z, f) = \mathcal{P}\left(\frac{\partial^4 N}{\partial M \partial q \partial z \partial \ln f_p} \Delta M \Delta q \Delta z \Delta \ln f\right)
\end{equation}
We generate multiple realizations by drawing many times from a Poisson distribution centered at that value. The frequency bins are chosen to align with the data preferences for various red-noise processes in the NANOGrav 15-year dataset \citep{2023ApJ...951L..10A}. These bins are centered at $f_i = i/T_{obs}$, where $i = 1, \ldots 14$ and $T_{obs} = 16.03$ yr \citep{2023ApJ...951L...8A}. 

At higher frequencies, binary systems emit GW with greater amplitude, but the noise power spectral density at these frequencies is also significantly higher. What's worse, these binaries evolve more rapidly, leading to fewer systems occupying each frequency bin. This reduction in the population outweighs the increase in amplitude due to the stronger power-law dependence. Therefore, our analysis can be safely restricted to $i\leq 14$. Additionally, we set 400 bins for $M$, logarithmically spaced between $M_{min}=10^7 M_\odot$ and $M_{max}=10^{11} M_\odot$. Then we set the size of redshift bins to be equal and given by $\Delta z = 0.1$ with $z_{min} = 0$ and $z_{max} = 5$. Similarly, the size of mass ratio bins is $\Delta q = 0.2$ with $q_{min} = 0.4$\footnote{Typical binary mass ratios are almost entirely above $q \sim 0.4$ as determined primarily by our choice of GMT $\propto q^{-1}$.} and $q_{max} = 1$. We vary the limits and sizes of the parameter bins and find only negligible changes.

\section{Gavitational Wave Background} \label{sec3}
\subsection{From Populations to GWB}
From the previous section, we have a population of SMBH binaries, which generate gravitational waves of different frequencies before their coalescence. The emitted energy per logarithmic frequency interval is given by $\mathrm{d} E_{\mathrm{gw}} / \mathrm{d} \ln f_{\mathrm{r}}$, where the energy is measured in the source rest frame at redshift $z$ and $f_{\mathrm{r}}=(1+z)f$ is the rest-frame frequency. 

The characteristic strain of the GWB produced by the superposition of the radiation from the binaries is therefore given by \cite{2001astro.ph..8028P}:
\begin{equation}
\begin{aligned}
    & \frac{\mathrm{d} \rho_{\mathrm{gw}}(f)}{\mathrm{d} \ln f} =\frac{\pi}{4} f^2 h_{\mathrm{c}}^2(f) \\
    & =\left.\int d M d q d z \frac{\partial^3 \eta}{\partial M \partial q \partial z} \frac{1}{1+z} \frac{\mathrm{d} E_{\mathrm{gw}}}{\mathrm{d} \ln f_{\mathrm{r}}}\right|_{f_{\mathrm{r}}=f(1+z)}.
\end{aligned}
\end{equation}
Here, we further assume all binaries are in circular orbits, emitting gravitational radiation predominantly at twice the orbital frequency, corresponding to the quadrupole order contribution. Then the energy emitted per logarithmic frequency interval is
\begin{equation}
\frac{\mathrm{d} E_{\mathrm{gw}}}{\mathrm{d} \ln f_{p}}= \frac{(G\mathcal{M})^{5 / 3}}{3G} (2 \pi f_{p})^{2 / 3}.
\end{equation}
Here, the observer-frame GW frequency $f$ relates to the rest-frame orbital frequency $f_p$ as $f=2 f_p /(1+z)$. The total binary mass $M$, and the chirp mass $\mathcal{M}$, are defined by
\begin{equation}
\mathcal{M}=\frac{\left(m_1 m_2\right)^{3 / 5}}{M^{1 / 5}}=M \frac{q^{3 / 5}}{(1+q)^{6 / 5}},
\end{equation}
with $q \equiv m_2 / m_1 (\leq 1)$ representing the binary mass ratio. Here we assume the mass ratio between two SMBHs is not extreme for the validity of Newtonian approximation. Combining the above equations we have
\begin{equation} \label{density_hc}
h_{\mathrm{c}}^2(f)= \frac{4 f^{-4 / 3}}{3 \pi^{1/3} c^2} \
\int d M d q d z \frac{\partial^3 \eta}{\partial M \partial q \partial z} \frac{(G \mathcal{M})^{5 / 3}}{(1+z)^{1 / 3}} ,
\end{equation}
where $G$ denotes the gravitational constant and $c$ is the speed of light. This relationship shows the connection between the merger number density and the GWB amplitude, leading to the widely used power-law expression for the GWB spectrum
\begin{equation} \label{pwer_form}
h_{\mathrm{c}}(f)=A_{\mathrm{yr}} \cdot\left(f / \mathrm{yr}^{-1}\right)^{\alpha},
\end{equation}
where the power law index $\alpha = -2/3$ for the standard GWB spectrum.
Combining Equations~(\ref{pop_from_density}), (\ref{pop_to_density}), (\ref{gw_time}) and~(\ref{density_hc}), we derive an alternative expression for the GWB's characteristic strain \citep{2008MNRAS.390..192S}
\begin{equation} \label{pop_hc}
h_{\mathrm{c}}^2(f)=\int d M d q d z \frac{\partial^4 N}{\partial M \partial q \partial z \partial \ln f_p} (1 + z)^2 h_{\mathrm{s}}^2\left(f_p\right).
\end{equation}
Here $h_{\mathrm{s}}$ represents the sky- and polarization-averaged strain amplitude from a single inspiraling binary, expressed as:
\begin{equation} \label{gw_strain}
h_{\mathrm{s}}^2(f_p)=\frac{G}{c^3} \frac{L_{\mathrm{GW}}}{\left(2 \pi f_p\right)^2 d_L^2}=\frac{32}{5 c^8} \frac{(G \mathcal{M})^{10 / 3}}{d_L^2}\left(2 \pi f_p\right)^{4 / 3},
\end{equation}
where $d_L$ is the luminosity distance to a source at redshift $z$.

Finally, after discretizing the population of SMBHBs by Equation~(\ref{pop_gen}), the GWB can be calculated from the discrete population by
\begin{equation} \label{pop_gw_gen}
h_{\mathrm{c}}^2(f)=\sum_{M, q, z, f} N_\mathcal{P}(M, q, z, f)  \frac{h_{\mathrm{s}}^2\left(f_p\right)}{\Delta \ln f} (1+z)^2.
\end{equation}

Compared to Equation~(\ref{density_hc}), we directly calculate the GWB spectrum as the cumulative GW emission from individual binaries across the Universe in Equation~(\ref{pop_gw_gen}). As a result, the GWB deviates from a single power-law spectrum.

\subsection{Late Merger vs Early Merger Model} \label{late_early}
We generate an ensemble of $10^4$ realizations for the population of SMBHBs for late merger and early merger models.  This number of realizations is sufficient to account for the Poissonian noise of rare binaries after checking the outcomes from $5000$ realizations. Then we calculate the GWB for each realization under different population models.

\begin{figure}
    \includegraphics[width=\linewidth]{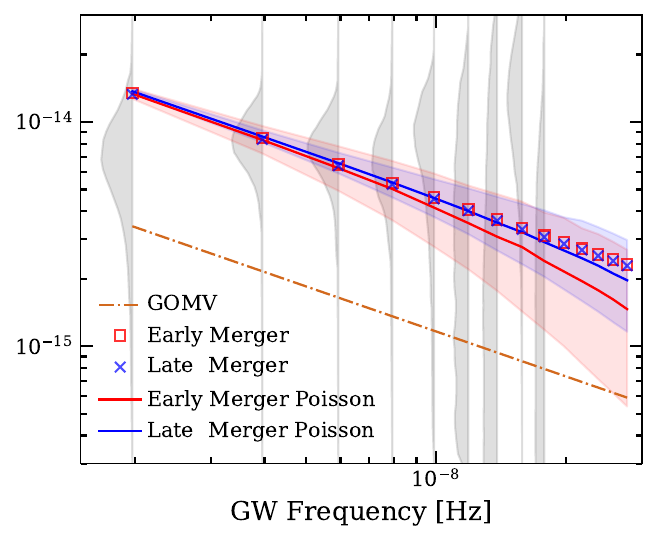}
    \caption{Characteristic GW strains, $\mathrm{h_{c}}$, as predicted by the Galaxy Observation Mean Value (GOMV; \customcite{2014ApJ...783...85T}), late merger, and early merger models. Grey violins illustrate the posteriors of $\mathrm{h_{c}}$ across frequency bins based on the HD-w/MP+DP+CURN model in \cite{2023ApJ...951L...8A}. Red squares and blue crosses represent the fitted power-law spectra ($\mathrm{h_{c}} \propto f^{-2/3}$) for early and late merger models based on the median of the posteriors. Solid lines indicate the ensemble average of $\mathrm{h_{c}}$ from $10^4$ realizations for each model. Shaded areas represent the range from the 10th to 90th percentiles, highlighting the variance due to Poissonian noise at higher frequencies where binaries are rarer, deviating from the ideal power-law.}
    \label{fig:h_c}
\end{figure}

\begin{figure*}
    \centering
    \includegraphics[width=1.\textwidth]{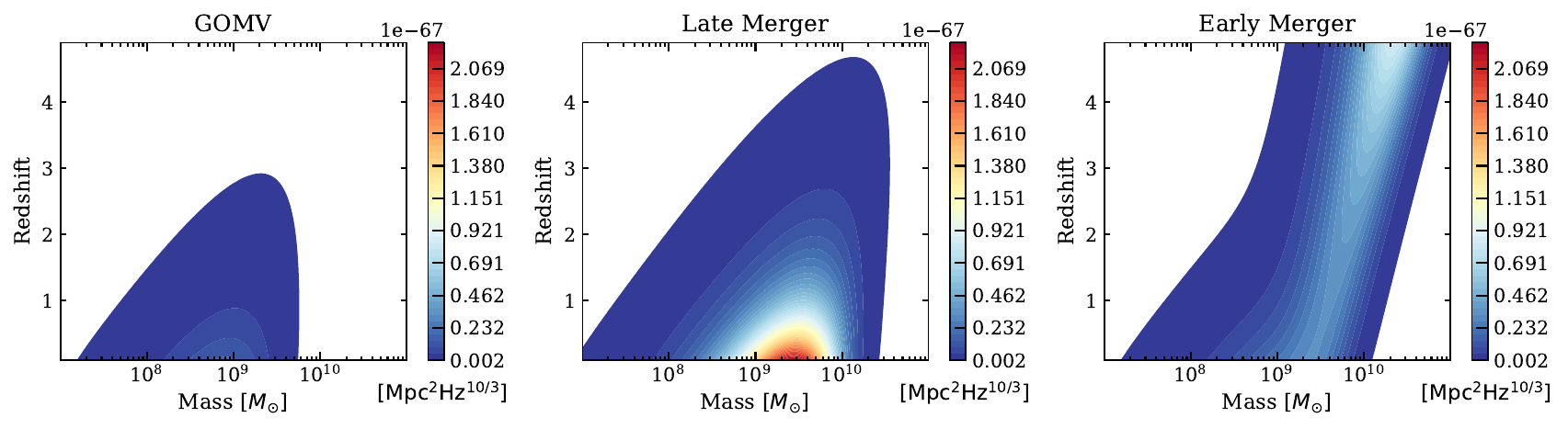}
    \caption{Differential contributions $\mathrm{h_{d}^2}$ to the characteristic gravitational wave strains $\mathrm{h_{c}^2}$ with $q = 1$, for the GOMV, late merger, and early merger models. Integrating the areas under the contours within each plot yields the total amplitude of $\mathrm{h_{c}^2}$ for the respective models. The GOMV model has a significantly lower amplitude of $\mathrm{h_{c}}$ in comparison to both the late and early merger models, which generate the similar amplitude of GWB. The late merger model primarily receives its contribution from low-redshift mergers with masses in the range of $10^9$ to $10^{10} M_\odot$. In contrast, the early merger model is distinguished by its substantial contributions from high-redshift mergers, especially those with masses greater than $10^{10} M_\odot$. Furthermore, the distribution peak for the late merger model is more concentrated, while the early merger model displays a broader peak, reflecting a more diverse range of contributing redshifts and masses.}
    \label{fig:h_c_diff}
\end{figure*}

The results are shown in Fig~\ref{fig:h_c}, where we compare the characteristic gravitational wave strains, $h_c$, as predicted by GOMV, late merger, and early merger models. Grey violin plots illustrate the posterior distributions of $h_c$ across frequency bins, derived from an analysis that includes HD-correlated noise and fits for multiple noise spectra (HD-w/MP+DP+CURN; \customcite{2023ApJ...951L...8A}), with only the first nine posteriors displayed for clarity. 

Henceforth, we focus on the NANOGrav results, given that EPTA+InPTA and PPTA have detected similar GWB amplitudes, and our subsequent analysis uses pulsar data from NANOGrav. It is obvious from Fig~\ref{fig:h_c} that the characteristic strain generated by the GOMV (brown dash line) is significantly lower than the observed GWB. The red squares and blue crosses represent the best-fitted power-law spectra for the late and the early merger models based on the median amplitudes of the posteriors from the HD-w/MP+DP+CURN model in \cite{2023ApJ...951L...8A}. Compared to the late merger model, the GOMV model features smaller normalizations of the GSMF and the scaling relation, as well as a lower turnover mass for the GSMF. These characteristics result in SMBHB mergers with lower mass and fewer occurrences in the GOMV model. Consequently, the $h_c$ produced by the GOMV model is significantly smaller than that of the late merger model. The early merger model, however, includes an additional population of galaxies and therefore an additional population of SMBHB mergers, leading to a higher GWB compared to the GOMV model.

The solid lines show the ensemble average of $h_c$ from the discrete populations. The shaded blue and red areas represent the range from the 10th to 90th percentiles, illustrating the Poissonian noise coming from discretization, which is bigger for binaries with higher frequencies as they are rarer. As expected, the continuous distributions tend to overestimate the amplitude of $h_c$, and the Poissonian noise results in a steeper spectrum from the ideal power-law. Although late and early merger models produce almost identical power-law spectra, their ensemble averages and variances due to Poissonian noise are different. The early merger model suffers more from discreteness than the late merger model, especially at high frequencies. 

While a given overall amplitude of the power-law GWB can be produced by either a larger number of lower-mass SMBH binaries or a smaller number of higher-mass binaries, these differences change the frequency at which discreteness becomes important. As a result, they change the location and severity of the high-frequency spectral steepening. 

Therefore, finding out the contribution of the GWB helps understand why the early merger model suffers more from discreteness. Let us define the differential contribution by 
\begin{equation} \label{hc_differential}
h_{\mathrm{d}}^2(M,q,z) = \ln(10) M \frac{\partial^3 \eta}{\partial M \partial q \partial z} \frac{(G \mathcal{M})^{5 / 3}}{(1+z)^{1 / 3}}.
 \end{equation}
Then Figure~\ref{fig:h_c_diff} shows $\mathrm{h_{d}^2}$ for the GOMV, late merger, and early merger models with $q = 1$. We multiply the integrand in Equation~(\ref{density_hc}) by $\ln(10)M$ to define $\mathrm{h_{d}^2}$ since $M$ is in logarithm scale in Figure~\ref{fig:h_c_diff}. Then the area of the contour times its value of $\mathrm{h_{d}^2}$ yield directly the contribution to the total amplitude of $\mathrm{h_{c}^2}$.

From the contours in Figure~\ref{fig:h_c_diff}, it is evident that the late merger model receives most of the contribution from lower-redshift binaries with masses in the range of $10^9$ to $10^{10} M_\odot$. The differential contribution of the late merger model has a concentrated peak and the overall shape is the same as the GOMV - this is because they both originate from a single-Schechter function. However, the GOMV model has a much smaller overall amplitude compared to the late merger model, and the peak of the late merger model is shifted towards more massive mergers. These features are expected as the late merger model has a higher normalization factor and turnover mass in the GSMF.

In contrast, the early merger model has substantial contributions from higher-redshift mergers, especially those with masses greater than $10^{10} M_\odot$. Furthermore, the distribution peak of the early merger model is much broader, showing a more diverse range of contributing redshifts and masses. These features originate from the more complex double-Schechter function of the GSMF for the early merger model, and the additional component contributes significantly.

We can now explain why the early merger model is more susceptible to discreteness issues. This primarily stems from differences in the hardening timescale and luminosity of the major contributing binaries. Given that the luminosity of a binary scales significantly with mass, the same amplitude in $h_\mathrm{d}^2$ corresponds to a relatively smaller population for massive mergers.

Furthermore, as indicated by Equation~(\ref{gw_time}), the hardening timescale is considerably shorter for more massive binaries. Additionally, binaries at high redshift exhibit shorter hardening timescales for a given frequency and mass, attributed to their higher rest frame frequencies.

Notice that the population proportionally correlates with its hardening time. Considering these factors, the population size of the additional high-redshift massive binaries is substantially limited. Therefore, the early merger models are expected to suffer more from the Poissonian noise. 

Current constraints on the GWB are limited. Only the first-order feature, represented by a simple power-law, has been characterized with considerable uncertainty. As observational data improves over time, higher-order features of the GWB spectrum, including the effects of Poissonian noise, will be gradually identified, providing more constraints on SMBH populations.

\section{First detectable single source} \label{sec4}
Given the potential for different SMBHB redshift evolutions to generate similar GWBs, the characteristics of the first detectable single source are pivotal for distinguishing between different scenarios. In this section, we evaluate the detection probabilities of single binaries for the early and late merger models and present the properties of the first detectable single source. We identify the source as the binary within a given realization that exhibits the highest signal-to-noise ratio (SNR).

\subsection{Spectral Noise Density}
To evaluate the probability of detecting single sources, it is necessary first to understand the noise characteristics of the detector. This study focuses on pulsar data from the NANOGrav 15-year dataset, utilizing the effective strain noise power spectral density ($S_{\mathrm{eff}}$), which incorporates all aspects of detector noise characterization into a single figure of merit.

Using the Python package \textsc{hasasia} \citep{2019JOSS....4.1775H}, based on methodologies developed in \cite{2019PhRvD.100j4028H}, we calculate $S_{\mathrm{eff}}$ for detecting a deterministic GW
source averaged over its initial phase, inclination, and sky
location across various PTA configurations and observation durations. This combination is based on the SNR of the GWB optimal statistic developed in \cite{2009PhRvD..79h4030A}, \cite{PhysRevD.91.044048}, and \cite{2015MNRAS.451.2417R}. To facilitate a focused comparison of the potential detection capabilities for single binary systems and their properties, only the ``Status quo" and ``Optimal" PTA configurations will be presented in subsequent analyses. 

Under the ``Status quo" configuration, no enhancements are made to the current PTA, such as increasing the number of pulsars or reducing the white noise level in each pulsar. Instead, we extend the observational period using real data from the 67 existing pulsars in the current NANOGrav 15-year PTA, which includes specific sky locations, total observation times, and white noise levels as outlined by \cite{2023ApJ...951L..10A}. The observational cadence is maintained at once every two weeks for each pulsar.

Additionally, we incorporate the red noise attributed to the stochastic GWB, designated as $A = 2.15\times 10^{-15}$ and $\alpha = -2/3$, based on Equation~(\ref{pwer_form}). For pulsars exhibiting significant red noise, characterized by a Bayes factor $\mathcal{B}>100$ using the Savage-Dickey approximation \citep{10.1214/aoms/1177693507}, their specific red noise power spectrum values are used instead. Detailed parameters for pulsars with significant red noise are provided in Table~\ref{tab:noise_params}. A comprehensive strategy to combine individual and GWB red noise contributions should be explored in future studies.

In the ``Optimal" configurations, we assume enhancements to the PTA by increasing the number of pulsars to $N_s = 150$. Each newly added pulsar is expected to achieve a timing accuracy of $\sigma_{new} = 0.2\mu s$ in the white noise level, with none exhibiting significant individual red noise detections. These additions are assumed to be included starting from the end of the NANOGrav 15-year observation. 

\begin{figure}
    \centering
    \includegraphics[width=\linewidth]{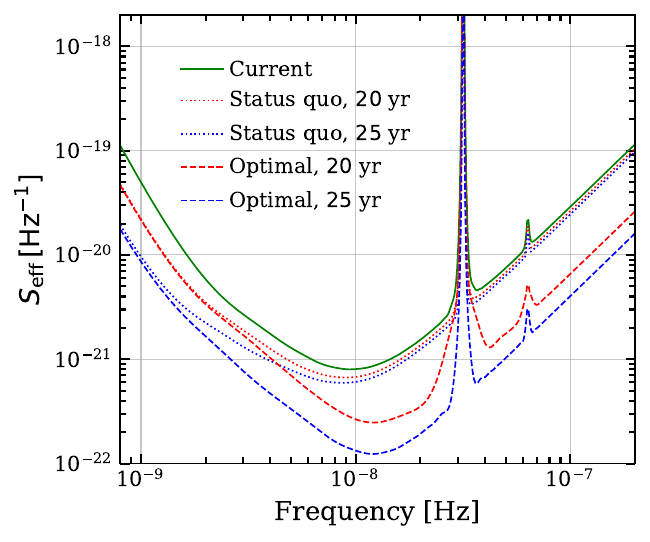}
    \caption{Effective strain noise power spectral density ($S_{\mathrm{eff}}$) of a deterministic source averaged over its initial phase, inclination, and sky location. In the ``Status quo" scenario, the existing PTA configuration is maintained without any additions, with noise reduction achieved solely by extending the observational period. Conversely, the ``Optimal" scenario implements a PTA configuration with $N_s = 150$ pulsars, achieving a timing accuracy of $\sigma_{new} = 0.2\mu s$ for the newly added pulsars. $S_{\mathrm{eff}}$ for both $T_\mathrm{span} = 20$ yr and $T_\mathrm{span} = 25$ yr are presented.}
    \label{fig:Future_Seff_Op}
\end{figure}

Figure~\ref{fig:Future_Seff_Op} presents the effective strain noise power spectral density averaged over inclination and sky location, $S_{\mathrm{eff}}$, for both the ``Status quo" and ``Optimal" scenarios. Notably, the increase in observation time significantly reduces $S_{\mathrm{eff}}$, particularly at low frequencies where red noise is predominant. This effect is less pronounced at higher frequencies, which are primarily influenced by the white noise associated with the pulsars' time of arrival.

It is noteworthy that $S_{\mathrm{eff}}$ in the white noise-dominated regime ($f \gtrsim 10^{-8}$ Hz) plays a vital role in the detection of individual SMBHBs. The GW strain from binaries at these frequencies is substantially greater than at lower frequencies for a fixed mass, and the evolution is not rapid enough to significantly diminish the probability of detection.
Incorporating more pulsars with precise timing into the array, as proposed in the optimal scenario, significantly enhances $S_{\mathrm{eff}}$ in regions dominated by white noise. This improvement occurs even though the newly added pulsars have not been observed for extended periods. 

In principle, the white noise of the pulsars in the current array should be improved not only through extended observation periods but also through advancements in radio telescope technologies and noise analysis techniques. However, to maintain a conservative approach, we exclude this potential improvement. We hope that these enhancements, combined with the addition of precisely timed pulsars, can be realized in the near future.

\subsection{Evolving Binary \& SNR}
As discussed at the end of Section~\ref{late_early}, SMBHBs that are both massive and highly redshifted evolve rapidly, especially at higher orbital frequencies. While many prior studies have treated SMBHBs as monochromatic sources \citep{2015MNRAS.451.2417R, 2023MNRAS.518.1802S}, this simplification may not hold, particularly under the early merger model considered in our analysis.

To address this, we evaluate the time derivative of the observed GW frequencies from SMBHBs, $df/dt$, using the relationship outlined in Equation~(\ref{gw_time}):
\begin{equation}  \label{df/dt}
\frac{\partial f}{\partial t}=\frac{384}{5}\left(\frac{G \mathcal{M}}{c^3}\right)^{5/3} \pi^{8/3} f^{5/3} (1+z)^{2/3}.
\end{equation}
Here, $T_\mathrm{span}$ represents the total observational time of the PTA, correlating to the lowest distinguishable frequency, $\sim 1/T_\mathrm{span}$. The total frequency change over the observation period can be approximated by $df/dt \cdot T_\mathrm{span}$. If this product exceeds $1/T_\mathrm{span}$, then the frequency shift is theoretically resolvable by the PTA.

Motivated by the above reasoning, we introduce the dimensionless rate of change factor, $\Delta\mathscr{F}$:
\begin{equation} \label{Delta_F}
\Delta\mathscr{F} = \frac{df}{dt} \times T_\mathrm{span}^2.
\end{equation}
A $\Delta\mathscr{F}$ value greater than one indicates that the binary is ``fast-evolving", making it likely that its frequency change is detectable. For our analyses, we distinguish these binaries from monochromatic GW sources when calculating the SNR, applying the monochromatic approximation only to binaries for which $\Delta\mathscr{F} < 1$.

To proceed further, we consider gravitational waves from a single binary system, averaged over its initial phase, inclination, and sky location. For a monochromatic source, the SNR is expressed as \cite{2019PhRvD.100j4028H}:
\begin{equation}
\label{mono_snr}
\rho \equiv \sqrt{\left\langle\rho^2\right\rangle_{\text {inc, sky }}} \simeq h_0 \sqrt{\frac{T_{\mathrm{span}}}{S_{\mathrm{eff}}\left(f_0\right)}}.
\end{equation}
Here, $T_{\mathrm{span}}$ represents the total observational time, and $f_0$ the frequency of the binary. The dimensionless amplitude of the binary $h_0$ is determined by\footnote{$S_{\mathrm{eff}}$ has accounted for the average of inclination and sky location.}:
\begin{equation}
h_0^2(\mathcal{M},z,f) = \frac{16}{c^8} \frac{(G \mathcal{M})^{10 / 3}}{d_L^2}\left( \pi f (1+z) \right)^{4 / 3}.
\end{equation}

For the fast-evolving binaries, the frequency evolution timescale is still small compared to the orbital period, we approximate the GW strain amplitude in the time domain as:
\begin{equation}
h(t) \simeq h_0(\mathcal{M},z,f(t))\cdot sin(2\pi tf(t)),
\end{equation}
with $f(t)$ derived from solving Equation~(\ref{df/dt}). The SNR for these rapidly evolving binaries is then calculated as\footnote{The stationary phase approximation is not applicable in this scenario as the binaries experiencing rapid evolution have entered the final stages of their inspiral phase. However, the evolution timescale for binaries detectable by PTAs is still long enough to prevent mergers within the observation period.}:
\begin{equation}
\rho^2 = \int_{f_1}^{f_{14}} \mathrm{d} f \frac{4|\tilde{h}(f)|^2}{S_{\mathrm{eff}}(f)}
\end{equation}
where $\tilde{h}(f)$ is the Fourier transform of the GW strain amplitude $h(t)$. The setup of the frequency bins, $f_i = i/T_{obs}$, is the same as before. 

\begin{figure}
    \centering
    \includegraphics[width=\linewidth]{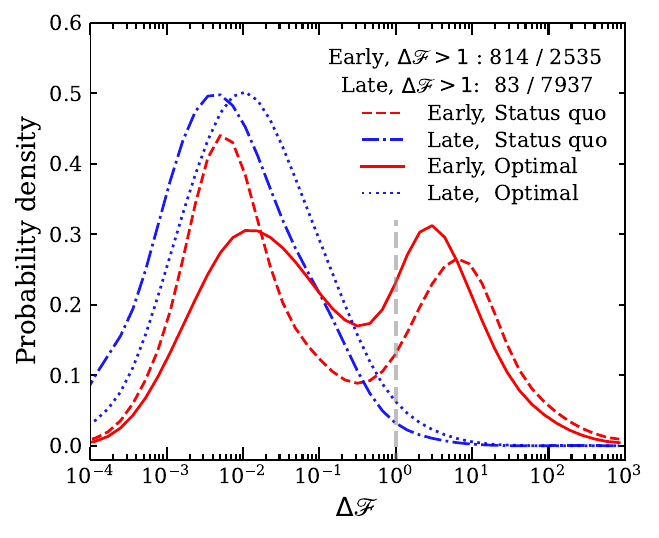}
    \caption{Probability distribution of the rate of change factor ($\Delta \mathscr{F}$) for the loudest binary in each realization. The logarithmic scale has been properly incorporated to ensure the area under the curve represents the correct fraction. The late merger model predominantly exhibits a peak at $\Delta \mathscr{F} \approx 10^{-2}$, whereas the early merger model displays dual peaks: one aligned with the late merger model and another at a significantly higher rate. The legend also shows the number of binaries with $\Delta \mathscr{F} > 1$ under the optimal configuration. Despite the late merger model hosting a larger number of realizations with SNR $>$ 3, it features significantly fewer rapidly evolving binaries ($\Delta \mathscr{F} > 1$) compared to the early merger model, where approximately 30\% of the loudest binaries are fast-evolving.}
    \label{fig:Delta_F}
\end{figure}

We can now calculate the SNR for all binaries and determine whether the loudest binary within each realization is fast-evolving. Figure~\ref{fig:Delta_F} shows the rate of change factor, $\Delta \mathscr{F}$, for the loudest binary in each realization under different merger models. The logarithmic scale has been properly accounted for and the total observation period is $T_\mathrm{span} = 25$ years. Only binaries with an SNR $>$ 3 are displayed, as this is considered the minimum detectable SNR, excluding realizations unlikely to detect a single binary within 10 years.

As shown in the figure, the late merger model exhibits a single dominant peak around $\Delta\mathscr{F} \approx 10^{-2}$. In contrast, the early merger model has two peaks: one similar to the late merger model and another at a significantly higher rate. The latter peak, with $\Delta\mathscr{F} > 1$, indicates a considerably higher probability for the first detected binary in the early merger model to exhibit detectable frequency evolution.

Interestingly, although the late merger model has about three times more realizations containing binaries with SNR $>$ 3, only about 1\% of these are fast-evolving. Conversely, the early merger model has fewer realizations with SNR $>$ 3 binaries, but approximately 30\% of these binaries are fast-evolving. The distribution characteristics are insensitive to PTA configurations.

This suggests that detecting the frequency evolution of the first single source would strongly favor the early merger model over the standard late merger model. While this conclusion is robust across different PTA configurations, the ability to detect such frequency evolution is contingent upon the specific PTA setup. Moreover, as the observational time increases, the ability to resolve finer frequency evolution improves, as indicated by Equation~(\ref{Delta_F}). Future studies should carefully account for these factors to ensure that any non-detection of frequency evolution is due to the intrinsic properties of the binaries.

\subsection{Detection Probability of a Single Binary}
As indicated by \cite{2015MNRAS.451.2417R}, simply assuming detection when the SNR achieves a threshold is inappropriate. A more accurate method involves using a matched filter technique to search for deterministic signals with unknown parameters.

For binaries with orbital evolution time-scales shorter than the typical pulsar-Earth light travel time but not detectable during PTA observation, the relevant parameter space for signal template construction can be reduced to three intrinsic parameters: frequency $f$ and sky location $\theta, \phi$ \citep{2012ApJ...756..175E,2012PhRvD..85d4034B}. Extrinsic parameters such as inclination and polarization angles are not formally searched over as they are highly degenerate.

For fast-evolving binaries, an additional intrinsic parameter should account for frequency evolution. However, only very massive high-frequency binaries, which are rare, require this consideration. These binaries are quasi-monochromatic, so a small extension of the templates should suffice.

Determining the optimal number of templates is complex and beyond this paper's scope. We use the same number of templates as in \cite{2015MNRAS.451.2417R}, considering only monochromatic sources, with a total template count of $N_t = 10^4$. This choice, although somewhat arbitrary, has minimal impact on our results after careful checking.

Once the signal templates are ready, we can use the $\mathcal{F}$-statistic to calculate the detection probability of single sources. The $\mathcal{F}$-statistic is optimal in the Neyman-Pearson sense \citep{2012LRR....15....4J, 2012ApJ...756..175E, 2012PhRvD..85d4034B}. 

\begin{figure}
    \centering
    \includegraphics[width=\linewidth]{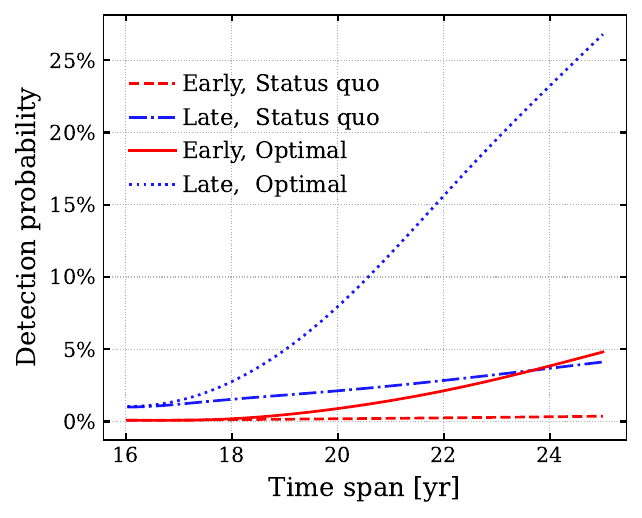}
    \caption{Ensemble average detection probability of single binaries as a function of observing time, comparing the status quo (dashed lines) and optimal (solid lines) configurations. The late merger model exhibits a significantly higher detection probability than the early merger model. Despite both models generating similar GWB, the substantial differences in detection rates of single binaries suggest that future observations could distinguish between them.}
    \label{fig:Detection_P}
\end{figure}

In the absence of a signal, the probability density function (PDF) of the $\mathcal{F}$-statistic follows a $\chi^2$ distribution with 4 degrees of freedom
\begin{equation}
p_0\left(\mathcal{F}\right)=\mathcal{F} e^{-\mathcal{F}}.
\end{equation}
When a signal is present, the PDF becomes a non-central $\chi^2$ distribution with 4 degrees of freedom
\begin{equation} \label{dp_r}
p_1\left(\mathcal{F}, \rho\right)=\frac{\left[2 \mathcal{F}\right]^{1 / 2}}{\rho} I_1\left(\rho \sqrt{2 \mathcal{F}}\right) e^{-\mathcal{F}-\frac{1}{2} \rho^2} .
\end{equation}
where $I_1(x)$ is the modified Bessel function of the first kind of order 1, and the non-centrality parameter $\rho$ is the SNR.

If all intrinsic parameters are known, the false alarm probability (FAP) for a single trial is given by
\begin{equation}
\alpha_s=\int_{\bar{\mathcal{F}}}^{\infty} p_0\left(\mathcal{F}\right) d \mathcal{F}=\left[1+\bar{\mathcal{F}}\right] e^{-\bar{\mathcal{F}}}.
\end{equation}
When parameters are unknown, the data must be filtered using a number of templates $N_t$ that cover the relevant parameter space of possible GW signals. Each template represents an independent trial, resulting in a total FAP of
\begin{equation}
\alpha=1-\left[1-\alpha_s\right]^{N_t}.
\end{equation}
By selecting a certain value of FAP, such as $\alpha = 1\%$ in this study, we can determine the threshold $\bar{\mathcal{F}}$. The detection probability is then calculated by integrating Equation~(\ref{dp_r}):
\begin{equation}
P_i =\int_{\bar{\mathcal{F}}}^{\infty} \frac{\left[2 \mathcal{F}\right]^{1 / 2}}{\rho} I_1\left(\rho \sqrt{2 \mathcal{F}}\right) e^{-\mathcal{F}-\frac{1}{2} \rho^2} d \mathcal{F}.
\end{equation}
This represents the probability of detecting one binary in a particular frequency bin. Therefore, the total probability of detecting at least one single source in any frequency bin is
\begin{equation}
P_{\mathrm{dec}}=1-\prod_i\left[1-P_i\right],
\end{equation}
where the index $i$ includes all frequency bins between $f_{\min}$ and $f_{\max}$ as detections in each bin are independent of each other.

Figure~\ref{fig:Detection_P} shows the ensemble average detection probability of single binaries as a function of observing time for both the status quo and optimal PTA configurations. Obviously, the late merger model exhibits a significantly higher detection probability than the early merger model, regardless of the PTA configuration.

Although both models generate similar GWBs, the substantial differences in the detection rates of single binaries indicate that they can be distinguished by future observations.

\subsection{Properties of the First Detected Source}
In this section, we select the SMBHB with the highest SNR in each realization and construct the distributions of their parameters. Unlike {\it LIGO-Virgo} binaries which can be used as standard sirens, it is impossible to estimate parameters such as the chirp mass and the luminosity distance for monochromatic binaries, as all parameters are degenerate into two parameters, a constant amplitude $h_0$ and an observed GW frequency. 

The observed frequency distribution is rather unremarkable, displaying a single prominent peak corresponding to the frequency with the lowest $S_{\mathrm{eff}}$. However, if the GWs from the binaries exhibit detectable frequency evolution, the degeneracy can be partially alleviated, potentially allowing inference of the mass and redshift of the detected binary. 

Here, we present the distributions of mass and redshift for the loudest SMBHB in each realization, excluding all realizations where the loudest SMBHB has an SNR less than three for $T_\mathrm{span} = 25$ years.

\begin{figure}
    \centering
    \includegraphics[width=\linewidth]{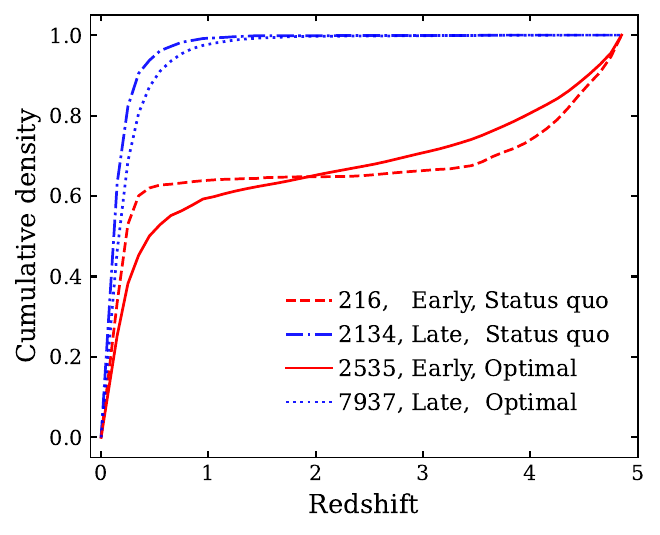}
    \caption{Cumulative redshift distribution of the loudest binaries for the late and early merger models. Only binaries with SNR $>$ 3 for $T_\mathrm{span} = 25$ years are presented, with the number of realizations containing such binaries shown in the legend. The late merger model predominantly features the loudest binaries originating from low redshifts ($z < 1$). In contrast, the early merger model demonstrates a balanced contribution from both low and high redshifts. The high-redshift component in the early merger model exhibits a broader extension, aligning with the extended peak observed in Figure~\ref{fig:h_c}.}
    \label{fig:z_distribution}
\end{figure}

Figure~\ref{fig:z_distribution} shows the cumulative redshift distribution of the loudest binaries for the late and early merger models. The late merger model predominantly features loud binaries from low redshifts ($z < 1$), whereas the early merger model shows a balanced contribution from both low and high redshifts, with a broader extension for the high-redshift component. This broadening aligns with the extended peak observed in Figure~\ref{fig:h_c}. Furthermore, these distribution features are insensitive to the assumed PTA configuration.

\begin{figure}
    \centering
    \includegraphics[width=\linewidth]{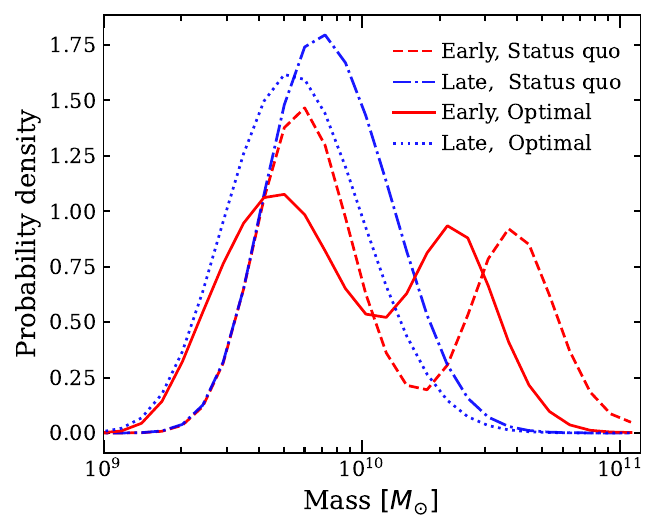}
    \caption{PDF of the total mass of the loudest binaries for both late and early merger models. The logarithmic scale has been properly accounted for to ensure the area under the PDF represents the correct fraction. Only binaries with SNR $>$ 3 for $T_\mathrm{span} = 25$ years are presented. Across all models and configurations, a pronounced peak is observed at $M \approx 5\times 10^9 M_\odot$. Notably, the early merger model exhibits an additional peak at higher total masses, originating from the high-redshift massive merger component.}
    \label{fig:m_distribution}
\end{figure}

Figure~\ref{fig:m_distribution} shows the PDF of the total mass of the loudest binaries. Across all models and configurations, a prominent peak appears at $M \approx 5 \times 10^9 M_\odot$, consistent with the masses of binaries that strongly dominate the GWB signal as reported by \cite{2023ApJ...952L..37A}. Similar to Figure~\ref{fig:Delta_F}, the early merger model exhibits an additional peak at higher total masses. The additional peak originates from the high-redshift massive merger component, and the fraction of very massive binaries ($M \gtrsim 10^{10} M_\odot$)  aligns with the fraction of high-redshift binaries in Figure~\ref{fig:z_distribution}.

Additionally, the status quo configuration exhibits a slight shift towards higher total masses and a significant reduction in the overall count of loud binaries. These features result from the elevated noise spectral density in the status quo configuration, which permits the detection of only very massive binaries from the high-redshift component. Nonetheless, the overall distribution features remain largely insensitive to the assumed PTA.

Overall, the features of these distributions follow our anticipations from the merger models and do not strongly depend on the assumed PTA configurations. Again, the ability to estimate these parameters should depend on the PTA configurations and is expected to suffer from significant degeneracy.

\section{Discussion} 
\subsection{Motivation of Early Merger Model} \label{motiva}

Even before the launch of JWST, substantial evidence suggested that galaxies could build up large stellar populations and quench star formation early \citep{2016A&A...592A.132G, 2020ApJ...892....7K, 2020ApJ...903...47F}. Observations have shown that the number density of massive galaxies exhibits little to no evolution from now to $z \sim 1.5$ \citep{2016A&A...592A.132G, 2020ApJ...892....7K} and changes very little up to $z \sim 4$ \citep{2015A&A...575A..96G, 2017A&A...605A..70D}.

Pushing this limit higher has been challenging because massive galaxies at high redshifts have low apparent magnitude and typically display very red colors, rendering them ``dark" to telescopes like the Hubble Space Telescope \citep{2024arXiv240308872W, 2024arXiv240408052B}. However, with JWST’s unprecedented resolution and sensitivity at longer wavelengths, we can achieve more accurate constraints on the abundance and masses of these galaxies at higher redshifts.

\cite{2024arXiv240308872W} demonstrates that UV-red galaxies dominate the high-mass end of the GSMFs at high redshifts. These UV-red galaxies have a much shallower power-law slope and form an additional bump at the high-mass end of the GSMF, similar to the additional population of massive galaxies shown in Figure~\ref{fig:GSMF_dec}. This bump gradually disappears at lower redshifts. Furthermore, the high-mass end of the GSMF of UV-red galaxies shows little evolution from $z \sim 4$ to $z \sim 6$. 

Therefore, the early merger model qualitatively accounts for the population of UV-red galaxies, characterized by its lower normalization $\Psi_0$, shallower power-law slope $\alpha_\psi$, and higher turnover mass $M_\psi$, with little redshift evolution.\footnote{The low-mass end of the GSMF of UV-red galaxies does exhibit significant evolution \citep{2024arXiv240308872W}, but this paper focuses on SMBHs with $M_{BH} > 10^9 M_\odot$.} However, the GSMF at the high-mass end for high redshifts ($z > 3$) remains significantly uncertain due to limited number counts from both observational constraints and the rarity of massive galaxies at high redshifts \citep{2024arXiv240308872W}. This is why we do not focus on the specific parameter values used in the early merger model. In the future, more comprehensive and robust population studies are necessary to precisely model the population of these massive early galaxies and to confirm that this enhanced population of massive high-redshift galaxies significantly impacts the GW signals detectable by future PTA observations.

Another component of the early merger model is the modified scaling relation between the mass of SMBHs and the stellar mass of their host galaxies. We introduce a strong redshift dependence on the normalization $\mu$, with higher redshifts tending to have more massive SMBHs. This choice is motivated by several observational results in the following.

Even before the launch of JWST, evidence suggested that black holes were overly massive compared to their host galaxies at high redshifts \citep{2006ApJ...649..616P, 2010ApJ...709..937G}, although this was limited to the most massive systems ($M_{\mathrm{BH}} > 10^8 M_\odot$). This counterintuitive result stimulated significant discussion regarding the co-evolution of black holes and bulges, and potential biases in measurement techniques.

However, the interpretation of these observations is fraught with uncertainty. SMBH mass estimates at cosmological distances are primarily obtained using luminous quasars, making it challenging to characterize the host galaxy properties when the quasar outshines the galaxy starlight by factors of  $\sim$ 10–30 \citep{2010ApJ...709..937G}. The launch of JWST has transformed this situation. With JWST's enhanced capabilities, \cite{2024ApJ...966..176Y} successfully disentangled the light from quasars and their surrounding stars, confirming that the most massive SMBHs are significantly overmassive compared to those in the local Universe.

Moreover, JWST is expanding our understanding of the high-redshift scaling relation by probing further and into smaller masses. Recent observations suggest that even less massive SMBHs at high redshift are overmassive compared to local galaxies \citep{2023ApJ...957L...3P, 2024arXiv240405793M, 2024arXiv240303872J}. Based on these observations, we propose a strong redshift dependence in the scaling relation for the additional component of the early merger model. We calibrated the normalization $\mu$ at $z = 5$ based on the scaling relation derived from AGNs observed at \(z = 4 \text{--} 7\) by JWST \citep{2023ApJ...957L...3P}, and set the local value of $\mu$ according to \cite{2013ARA&A..51..511K} and \cite{2013ApJ...764..184M}. We then assumed a linear increase in $\mu$ with redshift. Our scaling relation is consistent with the correlations derived from less massive SMBHs in other studies \citep{2023arXiv230801230M,2023ApJ...959...39H,2024arXiv240405793M}, though these correlations are not yet tightly constrained.

This modification could, in principle, apply to the normal component as well, but we adopt a conservative approach in light of studies suggesting no significant evolution in the scaling relation \citep{2023Natur.621...51D}. 
After all, the stellar masses of high-redshift quasar host galaxies remain highly uncertain \citep{2024ApJ...966..176Y, 2024arXiv240405793M}. Given the rapidly evolving research field, we qualitatively reproduce the observed trend and defer detailed modeling to future studies.

\subsection{Validity of Assumptions} \label{validity}
Throughout this paper, we assume all binaries are in circular orbits, and the decay of the orbit is purely due to the emission of GW. Nonetheless, real-world conditions may lead to deviations from these assumptions. However, these assumptions are good enough for this study, as argued in this section. 

The decay of orbits can originate from the interactions between the binaries and their environments including gravitational scattering by stars and interactions with circumbinary disks. These interactions can significantly accelerate the frequency evolution of the binary compared to the GW-only evolution, especially in the low-frequency regime $\left(f \ll 1 \mathrm{yr}^{-1}\right)$ where binaries can more easily couple to their local environments and GW emission is weaker. In fact, these interactions are necessary for the binaries to merge within a Hubble time. Therefore, there should be a flattening or turnover of the low-frequency GWB spectrum relative to the GW-only evolution due to energy transferred into the environments \citep{2011MNRAS.411.1467K}.  Consequently, environmental mechanisms are anticipated to play an important role in hardening the binaries sufficiently before GW emission becomes dominant. 

To quantify the effect of environmental interactions, we use the principle that the number of binaries entering and leaving each frequency bin at a given time should equal the merger rate. Therefore,
\begin{equation} \label{pop_time}
N(M,q,z,f) \propto \eta(M,q,z) \tau(M,q,z,f),
\end{equation}
where $N(M,q,z,f)$ represents the population of binaries, $\eta(M,q,z)$ is the merger rate, and $\tau(M,q,z,f)$ is the hardening timescale. This highlights that the frequency bin population proportionally correlates with the binary's hardening time, assuming a constant merger rate. Since different mechanisms of interactions will offer different hardening times, the population of SMBHBs and the corresponding GWB spectrum will change accordingly. 

However, we expect the first detected single binary by PTAs to have already entered the GW-dominated phase, given the enhanced strain from high-frequency binaries and the typical minimum of PTA noise spectral density. If the SMBH merger density, correlated with galaxy merger rates, is held constant, then the high-frequency binary population primarily reflects the GW decay dynamics. 

This reasoning can also be reflected in \cite{2023ApJ...952L..37A}, where the best fitting results from GW-only and phenomenological evolution (contains environmental interactions) generate almost identical spectra at higher frequencies. Therefore, the late merger model used in \cite{2023ApJ...952L..37A} will prefer the same population for galaxy mergers and high-frequency binaries based on the currently detected GWB, regardless of the choice of environmental interactions.

Environmental interactions, in this context, serve to bridge binaries to the threshold of GW-driven inspiral, introducing a temporal delay between the mergers of SMBHs and their host galaxies, in addition to the dynamical friction timescale. We assume efficient environmental coupling and approximate the total lifetime of binaries using the dynamical friction timescale from \cite{2008gady.book.....B}. This timescale characterizes the time taken for SMBHs to reach the center of their already merged host galaxies. For a typical velocity dispersion of $\sigma=200 \mathrm{~km} / \mathrm{s}$ and an initial separation of $r_i=5 \mathrm{kpc}$ between the SMBHs and the center of the merged galaxy, the timescale is approximately $t_{\mathrm{dy}} \sim 0.1 \mathrm{Gyr}$ for SMBHs with a mass of $M \sim 10^{10} M_{\odot}$. Consequently, we approximate the lifetime of binaries to be $0.1 \mathrm{Gyr}$ for all mergers.

Additionally, if a binary has eccentricity, it distributes GW energy across multiple integer harmonics. This redistribution results in a shift of energy toward higher frequencies, manifesting as a low-frequency turnover, a high-frequency flattening of the spectrum, and an intermediate ``bump" \citep{2007PThPh.117..241E,2017MNRAS.470.1738C}. Furthermore, binary eccentricity reduces the hardening timescale, especially for low-frequency binaries \citep{2013CQGra..30v4014S,2024A&A...685A..94E}. However, the pronounced manifestation of these spectral features requires binaries to have very high eccentricities (e.g., $e \gtrsim 0.9$) while having very small separations, which is unlikely to happen \citep{2023ApJ...952L..37A}. Therefore, we can safely restrict our analysis to circular binaries only.

\section{Conclusions and Outlook} \label{sec:conclusion}
In light of recent observations, we propose the early merger model, where an additional component of the SMBHs merged at greater redshifts, compared to the standard scenario where almost all the mergers take place at low redshift. The two models produce similar GWBs but exhibit several distinct observational outcomes that can be distinguished in future observations.

(i) The early merger model is expected to suffer more from Poissonian noise, leading to a steeper and earlier decline in the high-frequency GWB spectrum. This is because the produced GWB is dominated by a smaller number of SMBHBs with much higher masses and redshifts.

(ii) The early merger model has a much lower detection probability of single binaries. This is primarily due to the additional population of mergers being typically at high redshifts, where the distance significantly suppresses the SNR of individual binaries.

(iii) The early merger model has a much higher likelihood of the first detected binary exhibiting detectable frequency evolution. As shown in Equation~\eqref{df/dt}, binaries with higher mass and redshift have greater frequency evolution.

(iv) If frequency evolution can be detected, the parameter degeneracy of the single binary can be partially alleviated.  If the inferred mass and redshift of the first detected binary are high ($z >1$ and $M > 10^{10} M_\odot$), this will further favor the early merger model.

As mentioned before, these distinct observational outcomes are insensitive to the assumed PTA, but the detectability of these features does depend on the specific PTA configuration. Currently, the International Pulsar Timing Array (IPTA) is working to integrate data across all pulsars in the IPTA network \citep{2024ApJ...966..105A}. This effort includes not only adding new pulsars but also combining data from all three PTAs where any given pulsar is timed by multiple PTAs. Therefore, it is promising that the sensitivity of future PTAs could surpass our optimal configuration, enabling the observation to distinguish between the two merger models in the near future.

Additionally, as the JWST continues to collect data on the high-redshift Universe, the uncertainty in the high-redshift GSMF at the high-mass end should significantly decrease \citep{2024arXiv240308872W, 2024arXiv240408052B}. Confirming an enhanced population of massive high-redshift galaxies and more overmassive SMBHs within them will strongly support the early merger model.

Moreover, JWST is detecting more high-redshift AGNs than previously expected \citep{2024ApJ...963..129M,2024ApJ...964...39G}. Given that galaxy mergers are commonly associated with AGN activity, this provides further evidence in favor of the early merger model, which future observations will need to confirm.

If the early merger model is favored over the standard late merger model by future observations, it will greatly enhance our understanding of galaxy formation and evolution in the early Universe. This implies that the evolution pathways of galaxies at the high-mass end of the GSMF might differ significantly from those of less massive galaxies.

However, a more detailed modeling of the early merger model is necessary. The parameters used in the GSMF should be refined systematically and effects that modify the received GW signals, such as environmental interactions \citep{2020ApJ...901...25D,2023MNRAS.522.2707S} and gravitational lensing \citep{2023MNRAS.522.4059W}, should be incorporated into future studies.

Aside from SMBHs and their host galaxies, PTAs can also be utilized to study the nature of dark matter \citep{2023ApJ...951L..11A}. Specifically, PTAs can enhance our understanding of self-interacting dark matter \citep{2023arXiv230616966H}, which influences dark matter distribution within galaxies \citep{2021MPLA...3630001N}, and probe models of ultralight dark matter \citep{2022PhRvD.106c5032K,2024epsc.confE.132N}.

Now, PTAs have opened up the era of low-frequency GW astronomy, while JWST has unveiled a new vista for the high-redshift Universe. Together, these advancements herald a new age of discovery, where future observations and refined models will continually enrich our understanding of the Universe's earliest epochs.

\vspace{5mm}

          
\section*{acknowledgments}

This work was supported by the NSF AST-2107802, AST-2107806 and AST-2308090 grants. This research has made use of NASA's Astrophysics Data System Bibliographic Services.

\begin{table*}
\centering
\caption{Summary of model components with their symbols and values. All parameter values of the GOMV and Late Merger models align with those in \cite{2023ApJ...952L..37A}. \label{tab:model_components}}
\begin{tabular}{ccccc}
\hline\hline
\textbf{Model Component} & \textbf{Symbol} & \textbf{GOMV} & \textbf{Late Merger} & \textbf{Early Merger, added} \\
\hline
GSMF$(\Psi)$ & 
$\begin{array}{c}
\psi_0 \\
\psi_z \\
m_{\psi, 0} \\
m_{\psi, z} \\
\alpha_{\psi, 0} \\
\alpha_{\psi, z}
\end{array}$ & 
$\begin{array}{c}
-2.56 \\
-0.60 \\
10.9 \\
+0.11 \\
-1.21 \\
-0.03
\end{array}$ & 
$\begin{array}{c}
-2.25 \\
\ldots \\
11.25 \\
\ldots \\
\ldots \\
\ldots
\end{array}$ & 
$\begin{array}{c}
-3.56 \\
-0.03 \\
11 \\
-0.01 \\
1 \\
0
\end{array}$ \\
\hline
GPF$(P)$ & 
$\begin{array}{l}
P_0 
\end{array}$ & 
$\begin{array}{c}
+0.033 
\end{array}$ & 
$\begin{array}{l}
\ldots 
\end{array}$ & 
$\begin{array}{c}
\ldots 
\end{array}$ \\
\hline
GMT$(T_{\text{gal-gal}})$ & 
$\begin{array}{l}
T_0 \\
\beta_{t} \\
\gamma_{t} 
\end{array}$ & 
$\begin{array}{c}
+0.5 \mathrm{Gyr} \\
-0.5 \\
-1.0 
\end{array}$ & 
$\begin{array}{l}
\ldots \\
\ldots \\
\cdots
\end{array}$ & 
$\begin{array}{c}
\ldots \\
\ldots \\
\ldots
\end{array}$ \\
\hline
$M_{\mathrm{BH}}-M_{\text{bulge}}(M_{\mathrm{BH}})$ & 
$\begin{array}{c}
\mu \\
\alpha_\mu \\
f_{\star, \text{bulge}}
\end{array}$ & 
$\begin{array}{c}
8.6 \\
1.2 \\
0.615
\end{array}$ & 
$\begin{array}{c}
8.7 \\
\ldots \\
\ldots
\end{array}$ & 
$\begin{array}{c}
8.6 + 0.2 z \\
\ldots \\
\ldots
\end{array}$ \\
\hline
Binary Lifetime $\left(\tau_{tot}\right)$ & 
$\begin{array}{c}
\tau_{tot} \\
\end{array}$ & 
$\begin{array}{c}
0.1 \mathrm{Gyr} \\
\end{array}$ & 
$\begin{array}{c}
\ldots \\
\end{array}$ & 
$\begin{array}{c}
\ldots \\
\end{array}$ \\
\hline\hline
\end{tabular}
\centering 
\end{table*}

\begin{deluxetable}{ccc}
\tablecolumns{3}
\tablewidth{0pt}
\tablecaption{Puslars with significant red noises. The parameters $A_{RN}$ and $\gamma_{RN}$ correpsond to the standard convention given in Equation~(\ref{pwer_form}) and $\gamma=3+2 \alpha$. Here we only use the mean of the values and do not take the errors into account. \label{tab:noise_params}}
\tablehead{
    \colhead{Pulsar} & \colhead{\hspace{13mm}$\log_{10} A_{\mathrm{RN}}$\hspace{14mm}} & \colhead{\hspace{2mm}$\gamma_{\mathrm{RN}}$\hspace{5mm}}
}
\startdata
${\mathrm{B 1855+09}}$ & \hspace{14mm}$-14.0_{-0.4}^{+0.3}$\hspace{14mm} & \hspace{14mm}$3.9_{-0.8}^{+1.0}$\hspace{14mm} \\
B1937+21 & \hspace{14mm}$-13.6_{-0.1}^{+0.1}$\hspace{14mm} & \hspace{14mm}$4.0_{-0.3}^{+0.4}$\hspace{14mm} \\
B1953+29 & \hspace{14mm}$-12.8_{-0.3}^{+0.2}$\hspace{14mm} & \hspace{14mm}$1.8_{-0.7}^{+1.1}$\hspace{14mm} \\
J0030+0451 & \hspace{14mm}$-14.4_{-0.5}^{+0.4}$\hspace{14mm} & \hspace{14mm}$4.6_{-0.9}^{+1.1}$\hspace{14mm} \\
J0437-4715 & \hspace{14mm}$-13.4_{-0.2}^{+0.2}$\hspace{14mm} & \hspace{14mm}$0.5_{-0.4}^{+0.6}$\hspace{14mm} \\
J0610-2100 & \hspace{14mm}$-12.9_{-0.5}^{+0.3}$\hspace{14mm} & \hspace{14mm}$4.1_{-1.9}^{+2.0}$\hspace{14mm} \\
J0613-0200 & \hspace{14mm}$-13.8_{-0.3}^{+0.3}$\hspace{14mm} & \hspace{14mm}$3.1_{-0.7}^{+0.9}$\hspace{14mm} \\
J1012+5307 & \hspace{14mm}$-12.6_{-0.1}^{+0.1}$\hspace{14mm} & \hspace{14mm}$0.8_{-0.3}^{+0.3}$\hspace{14mm} \\
J1600-3053 & \hspace{14mm}$-13.5_{-0.6}^{+0.2}$\hspace{14mm} & \hspace{14mm}$1.6_{-0.7}^{+1.5}$\hspace{14mm} \\
J1614-2230 & \hspace{14mm}$-14.9_{-0.8}^{+1.0}$\hspace{14mm} & \hspace{14mm}$4.7_{-2.0}^{+1.6}$\hspace{14mm} \\
J1643-1224 & \hspace{14mm}$-12.3_{-0.1}^{+0.1}$\hspace{14mm} & \hspace{14mm}$0.9_{-0.4}^{+0.4}$\hspace{14mm} \\
J1705-1903 & \hspace{14mm}$-12.6_{-0.1}^{+0.1}$\hspace{14mm} & \hspace{14mm}$0.5_{-0.3}^{+0.4}$\hspace{14mm} \\
J1713+0747 & \hspace{14mm}$-14.1_{-0.1}^{+0.1}$\hspace{14mm} & \hspace{14mm}$2.6_{-0.4}^{+0.5}$\hspace{14mm} \\
J1738+0333 & \hspace{14mm}$-14.6_{-0.6}^{+0.8}$\hspace{14mm} & \hspace{14mm}$5.2_{-1.8}^{+1.3}$\hspace{14mm} \\
J1744-1134 & \hspace{14mm}$-14.1_{-0.6}^{+0.4}$\hspace{14mm} & \hspace{14mm}$3.6_{-1.2}^{+1.8}$\hspace{14mm} \\
J1745+1017 & \hspace{14mm}$-11.9_{-0.1}^{+0.1}$\hspace{14mm} & \hspace{14mm}$2.4_{-0.5}^{+0.6}$\hspace{14mm} \\
J1747-4036 & \hspace{14mm}$-12.6_{-0.2}^{+0.1}$\hspace{14mm} & \hspace{14mm}$2.4_{-0.7}^{+1.0}$\hspace{14mm} \\
J1802-2124 & \hspace{14mm}$-12.2_{-0.2}^{+0.2}$\hspace{14mm} & \hspace{14mm}$1.8_{-0.6}^{+0.7}$\hspace{14mm} \\
J1853+1303 & \hspace{14mm}$-13.5_{-0.4}^{+0.2}$\hspace{14mm} & \hspace{14mm}$2.3_{-0.7}^{+1.1}$\hspace{14mm} \\
J1903+0327 & \hspace{14mm}$-12.2_{-0.1}^{+0.1}$\hspace{14mm} & \hspace{14mm}$1.5_{-0.4}^{+0.4}$\hspace{14mm} \\
J1909-3744 & \hspace{14mm}$-14.5_{-0.4}^{+0.3}$\hspace{14mm} & \hspace{14mm}$4.1_{-0.9}^{+1.0}$\hspace{14mm} \\
J1918-0642 & \hspace{14mm}$-13.8_{-0.7}^{+0.4}$\hspace{14mm} & \hspace{14mm}$2.7_{-1.0}^{+1.5}$\hspace{14mm} \\
J1946+3417 & \hspace{14mm}$-12.5_{-0.1}^{+0.1}$\hspace{14mm} & \hspace{14mm}$1.4_{-0.4}^{+0.5}$\hspace{14mm} \\
J2145-0750 & \hspace{14mm}$-12.9_{-0.1}^{+0.1}$\hspace{14mm} & \hspace{14mm}$0.6_{-0.4}^{+0.5}$\hspace{14mm} \\
J2234+0611 & \hspace{14mm}$-13.9_{-0.9}^{+0.5}$\hspace{14mm} & \hspace{14mm}$3.2_{-1.9}^{+2.7}$\hspace{14mm} \\
\enddata
\end{deluxetable}

\bibliography{Foison_repository}{}
\bibliographystyle{aasjournal}


\end{document}